\documentclass[times,preprint,3p,sort&compress]{elsarticle}  
\usepackage{amssymb}
\usepackage{amsmath}

\usepackage{graphicx}  
\usepackage{bm}        
\usepackage{amssymb}   
\usepackage{amsbsy}
\usepackage{amsmath}
\usepackage{amsfonts}
\usepackage{color}
\usepackage[dvipsnames]{xcolor}
\usepackage{amsthm}
\usepackage{graphics}
\usepackage{graphicx}
\usepackage{xcolor}
\usepackage{caption}
\usepackage{subcaption}
\usepackage{amssymb}
\usepackage{enumitem}
\setlist[itemize]{leftmargin=*}

\usepackage{comment}

\usepackage{xcolor}
\usepackage{graphicx}

\usepackage[hidelinks]{hyperref}
\hypersetup{
  colorlinks   = true, 	
  urlcolor     = blue, 	
  linkcolor    = blue, 	
  citecolor   = red 	
}

\begin{document}
\begin{frontmatter}

\title{Excited Abelian-Higgs vortices: decay rate and radiation emission}
\author[salamanca1,salamanca2]{A. Alonso-Izquierdo}\ead{alonsoiz@usal.es}

\author[bilbao1,bilbao2,bilbao3]{J. J. Blanco-Pillado}\ead{josejuan.blanco@ehu.es}
\author[valladolid]{D.~Migu\'elez-Caballero}
\ead{david.miguelez@uva.es}
\author[valladolid]{S.~Navarro-Obreg\'on}
\ead{sergio.navarro.obregon@uva.es}
\author[salamanca1,salamanca2]{J. Queiruga\corref{cor1}}
\ead{xose.queiruga@usal.es}

\cortext[cor1]{Corresponding author}

\address[salamanca1]{Departamento de Mat\'ematica Aplicada, Universidad de Salamanca, Casas del Parque 2, 37008, Salamanca, Spain}
\address[salamanca2]{IUFFyM, Universidad de Salamanca, Plaza de la Merced 1, 37008, Salamanca, Spain}

\address[bilbao1]{Department of Theoretical Physics, University of the Basque Country UPV/EHU}

\address[bilbao2]{EHU Quantum Center, University of the Basque Country, UPV/EHU}

\address[bilbao3]{
IKERBASQUE, Basque Foundation for Science, 48011, Bilbao, Spain}
\address[valladolid]{Departamento de F\'{\i}sica Te\'{o}rica, At\'{o}mica y \'{O}ptica,
Universidad de Valladolid, 47011 Valladolid, Spain}

\begin{abstract}

The evolution of 1-vortices when their massive bound mode is excited is investigated in detail (both analytically and numerically) in the Abelian-Higgs model for different ranges of the self-coupling constant. The dependence of the spectrum of the 1-vortex fluctuation operator on the model parameter is discussed initially. A perturbative approach is employed to study the radiation emission in both the scalar and the vector channels. 
Our findings reveal that the oscillating initial configuration of the 1-vortex radiates at a frequency twice that of the internal mode. Through energy conservation considerations, we derive the decay law of the massive mode.  Finally, these analytical results are compared with numerical simulations in field theory. 
\end{abstract}


\end{frontmatter}

\section{Introduction}\label{Sec:1}

Among the range of topological solitons, vortices have emerged as notably significant and versatile solutions. They find applications in diverse areas of physics, such as superconductivity \cite{Abrikosov1957} or superfluidity \cite{superfluid-vortex} in condensed matter or particle physics models in cosmology \cite{Nielsen1973,Vilenkin2000}.

The Abelian-Higgs model is the prototypical model supporting relativistic gauged vortices (see  \cite{Manton2004,Taubes1980} and references therein). This model, which describes the minimal coupling between a $U(1)$ gauge field and a charged scalar field in a phase where the gauge symmetry is broken spontaneously, has been thoroughly studied over the last decades, leading to a deeper comprehension of the phenomena associated with this class of topological solitons. Research has shed light on fundamental aspects of vortices \cite{Nielsen1973,Penin2020,Perivolaropoulos1993}, their behavior in scattering processes \cite{Samols1992,Burzlaff1996,Stuart1994}, or the application of collective coordinates to reduce the degrees of freedom of the system \cite{Speight1997}. Generalizations of the model have also been considered in the literature, including dielectric terms \cite{Lee1991, Ghosh1994, Fuertes1999, Fuertes2014, Lima2020, Bazeia2019, Bazeia2022, Lima2022, Alonso2022}, or magnetic impurities \cite{Cockburn2017, Tong2014}.

In recent years, there has been a growing interest in the investigation of the excitation modes of topological defects. The spectral structure arising from small fluctuations around Abelian-Higgs vortices has been a topic of discussion in the last years by several authors. In a seminal paper by Weinberg \cite{Weinberg1979}, it was demonstrated that, in the BPS limit, the vortices exhibit $2n$ zero fluctuation modes. The problem of describing massive bound modes was initially tackled by Arodz in \cite{Arodz1991}. Subsequently, Goodband and Hindmarsh numerically studied the spectral structure of a vortex by assuming different winding numbers \cite{Goodband1995}. The study of zero modes and positive bound modes has been further analyzed by Alonso-Izquierdo \textit{et al} in the BPS limit \cite{Izquierdo2016,AlonsoIzquierdo2016} employing a hidden supersymmetric structure to simplify the problem. Furthermore, this work has led recently to the study of the spectral problem associated to the second order fluctuation operator for a configuration where two 1-vortices are located at an arbitrary distance \cite{Izquierdo2024}.

These modes may become particularly relevant when studying the dynamics of vortices
and their $3+1$ dimensional extensions, cosmic strings. The excitation of these solitonic structures may alter the mechanical properties of the strings, changing their energy and/or their tension. This suggests that further investigation into the dynamics of realistic strings extending beyond the thin wall approximation may be
necessary in some scenarios.

In fact, this idea has been recently suggested in \cite{Hindmarsh:2021mnl} as a possible way to reconcile the numerical results obtained in field theory simulations of cosmic string networks \cite{Hindmarsh:2017qff} and their Nambu-Goto counterparts in \cite{Blanco-Pillado:2013qja}. An initial examination of the dynamics of several loops from field theory simulations does not appear to indicate important deviations from the Nambu-Goto dynamics \cite{Blanco-Pillado:2023sap}. Therefore, the significance of these excitation modes for the conclusions of field theory simulations is not presently clear.

However, before we can understand the cosmological implications of these excited modes on strings 
one needs to investigate their most basic properties including their stability and potential decay rates via radiation emission. Note that understanding these properties is critical before extrapolating our results from a necessary limited numerical simulations to a cosmological context. 

In this manuscript we will thoroughly describe the evolution of a 1-vortex when its internal mode has been initially excited for different values of the self-coupling constant. For this purpose, we will employ a perturbative approach similar to the one introduced by Manton and Merabet for the study of kink excitations \cite{Manton1997}. This analytical method has also been successfully employed to examine the evolution of kinks numerically \cite{Blanco-Pillado:2020smt, Navarro2023} as well as global vortices \cite{BlancoPillado2021,Blanco-Pillado:2022axf} and wobblers in two component scalar field theories \cite{AlonsoIzquierdo2023, AlonsoIzquierdo2024}. The resulting expressions of this perturbative analysis enable us to confirm that a 1-vortex radiates through scalar and gauge fields, with a predominant frequency which is twice that of the excited internal mode. It is clear that the coupling between the massive bound mode and the radiation modes will lead to the subsequent decay of the internal shape mode amplitude. Employing perturbation theory, we derive a temporal decay law for this mode which is well approximated by an inverse square root expression which is in very good agreement with our numerical simulations.

The backreaction of excitations of the longitudinal component of the $U(1)$ gauge field was studied in a 3+1 dimensional model in \cite{Arodz1996, Arodz1997}. Note that these are physically different modes than the ones studied here. In fact, they are not present in a $2+1$ dimensional vortex we study in the present paper. Furthermore, the mathematical approach and our results differ from the ones used in the aforementioned articles. We leave the discussion of these other modes for a future publication.

This manuscript is organized as follows: In Section \ref{Sec:2}, a brief overview of the internal mode structure of the Abelian-Higgs 1-vortex is provided. Section \ref{Sec:3} offers a detailed perturbative study of the evolution of a 1-vortex whose internal mode has been initially excited. In Section \ref{Sec:4}, the validity of these results will be compared with those obtained from numerical simulations. Finally, Section \ref{Sec:5} summarizes the main findings of this work and outlines potential future prospects.

\section{The Abelian-Higgs model: Vortex solution and internal structure}\label{Sec:2}

The Abelian-Higgs model describes the coupling between a $U(1)$ gauge field and a complex scalar field in $2+1$ dimensions. The Lagrangian density that governs the dynamics of this model reads
\begin{equation}  \label{Eq:LagrangianDensity_0}
    \tilde{\mathcal{L}}=-\frac{1}{4}F_{\mu\nu}F^{\mu\nu}+\frac{1}{2}(D_{\mu}\Phi)^* D^{\mu}\Phi-\frac{\tilde{\lambda}}{8}(\eta^2-\Phi\Phi^*)^2,
\end{equation}
where the complex scalar field, $\Phi(x)=\Phi_1(x)+ i \Phi_2(x)$, represents a Higgs field and $A_\mu(x)=\left(A_0(x),A_1(x),A_2(x)\right)$ is the vector potential. For later convenience we rescale the fields and the coordinates as follows
\begin{equation}
\Phi\rightarrow \eta\, \Phi,\quad A_\mu \rightarrow \eta \,A_\mu,\quad x^\mu \rightarrow 1/\left(e\eta\right)\, x^\mu.
\end{equation}

In terms of the rescaled fields the Lagragian density reads
\begin{equation} \label{Eq:LagrangianDensity}
    \mathcal{L}=e^2\eta^4\left(-\frac{1}{4}F_{\mu\nu}F^{\mu\nu}+\frac{1}{2}(D_{\mu}\Phi)^* D^{\mu}\Phi-\frac{\lambda}{8}(1-\Phi\Phi^*)^2\right).
\end{equation}
The parameter $\lambda$ measures the quotient between the penetration lengths of the scalar and electromagnetic fields
in a superconducting medium as $\lambda = \tilde{\lambda}/e^2$. In addition, it can be seen as the quotient between the masses of the Higgs particle, $m_H = \sqrt{\tilde{\lambda}} \,\eta$, and the vector meson, $m_V = e\,\eta$, with $\eta$ the vacuum expectation value. The coupling constant $\lambda$ determines the strength of the Higgs potential. In terms of the rescaled field, the covariant derivative is defined as $D_\mu=\partial_\mu- i A_\mu (x)$ , the electromagnetic tensor as $F_{\mu \nu}(x)=\partial_{\mu}A_\nu(x)-\partial_{\nu}A_{\mu}(x)$, and the Minkowski metric is chosen in the form $g_{\mu \nu}={\rm diag}\{1,-1,-1\}$.  In (\ref{Eq:LagrangianDensity}), $\Phi^*$ stands for the complex conjugate of $\Phi$. From now on, the temporal gauge $A_0=0$ will be imposed. Consequently, the field equations associated to \eqref{Eq:LagrangianDensity} will be expressed as follows:
\begin{equation}\label{Eq:EquationSecondOrdenCartesianCoordinates}
    \begin{split}
         \partial_{0,0}\Phi  &=D_1D_1\Phi+D_2D_2\Phi+\frac{\lambda}{2}(1-\Phi\Phi^*)\Phi,\\
       \partial_{0,0}A_1  &=\partial_{2,2}A_1-\partial_{1,2}A_2-\frac{i}{2}\left(\Phi^* D_1\Phi-\Phi (D_1\Phi)^*\right),\\
        \partial_{0,0}A_2 &=\partial_{1,1}A_2-\partial_{1,2}A_1-\frac{i}{2}\left(\Phi^* D_2\Phi-\Phi (D_2\Phi)^*\right).
    \end{split}
\end{equation}
 
\begin{figure}[ht!]
    \centering
    
    \begin{subfigure}{0.489\textwidth}
        \includegraphics[width=\linewidth]{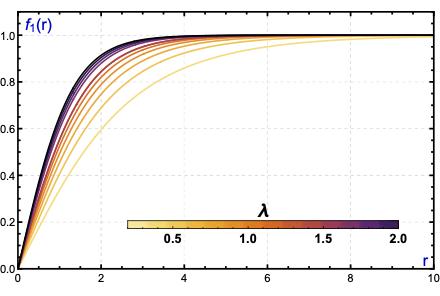}
    \end{subfigure}
    \hfill
    \begin{subfigure}{0.489\textwidth}
        \includegraphics[width=\linewidth]{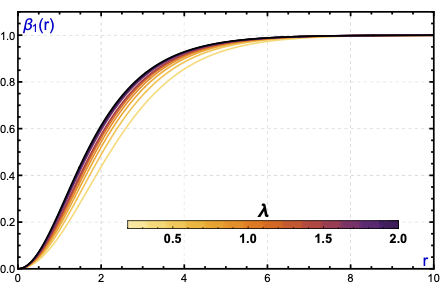}
    \end{subfigure}
  
    \caption{Scalar (left) and vector (right) component profiles of the 1-vortex for different values of the self-coupling constant $\lambda$.}
    \label{Fig:Profiles}
\end{figure}

For a static configuration the energy can be written as,
\begin{equation}
\label{eq:staticenergy}
    V=\frac{1}{2}\int_{\mathbb{R}^2}\left[ B^2+\left(D_1 \Phi \right)^*D_1 \Phi + \left(D_2 \Phi \right)^*D_2\Phi +\frac{\lambda}{4}(1-\Phi\Phi^*)^2 \right]dx_1 dx_2,
\end{equation}
where $B=F_{1 2}=\partial_1A_2-\partial_2 A_1$ is the magnetic field. In fact, (\ref{eq:staticenergy}) describes the free energy of the Ginzburg-Landau theory of superconductivity in the non-relativistic limit. In this scenario, $|\Phi(x)|^2$ represents the electron density of the physical substrate, while the value of the self-coupling constant $\lambda$ distinguishes if the material behaves as a Type I ($\lambda<1$) or a Type II ($\lambda>1$) superconductor. The transition point $\lambda=1$ corresponds to the so-called BPS or critical value, where the Higgs and gauge field masses are equal. Vortices are finite energy solutions of the field equations (\ref{Eq:EquationSecondOrdenCartesianCoordinates}). For this reason, they must satisfy the following asymptotic conditions on the circle at infinity $S_\infty^1$~,
\begin{equation}\label{Eq:BoundaryConditions}
   \Phi^* \Phi \, \big|_{S_\infty^1} = 1, \qquad D_i\Phi \big|_{S_\infty^1} = 0, \qquad B \big|_{S_\infty^1} = 0~.
\end{equation}
 By employing a polar coordinates system with $x_1= r \cos \theta$ and $x_2= r \sin \theta$ the conditions (\ref{Eq:BoundaryConditions}) suggest that $\Phi_\infty: =\lim_{r\rightarrow\infty}\Phi=e^{in\theta}$ with $n\in \mathbb{Z}$. This means that the vortices must asymptotically take values on a unit circle in field space. As a consequence, these solutions can be classified by the vorticity or winding number $n$ of the map $\Phi|_\infty : S_\infty^1 \rightarrow S^1$, which by topological considerations must be a conserved quantity of the system (see \cite{Manton2004,Speight1997}). On the other hand, the asymptotic behavior of the vector field $A$ is also fixed as 
\begin{equation}\label{Eq:A1A2Infinity}
    \left(A_1, A_2\right)=\left(-i e^{-i n \theta }\partial_1 e^{i n \theta },-i e^{-i n \theta }\partial_2 e^{i n \theta } \right),
\end{equation}
by the conditions (\ref{Eq:BoundaryConditions}). The direct consequence of this topological argument is that the magnetic flux defined by
\begin{equation}
    \Theta=\int_{\mathbb{R}^2}B\, dx_1 dx_2= \oint_{\mathbf{S}_\infty^1} \left(A_1 dx_1+A_2 dx_2\right)=2 \pi n ~,
\end{equation}
is classically quantized in the physical system. In order to find static circularly-symmetric $n$-vortex solutions, we use the following ansatz
\begin{equation}\label{Eq:AnsatzRadial}
    \Phi(r,\theta)=f_n(r) e^{i n \theta} \hspace{0.4cm} , \hspace{0.4cm}  A_r(r,\theta)=0  \hspace{0.4cm} ,  \hspace{0.4cm} A_\theta(r,\theta)= \frac{n\, \beta_{n}(r)}{r}  \hspace{0.2cm} ,
\end{equation}
where we have used the relations $A_r=A_1 \cos\,\theta + A_2\sin\, \theta$ and $A_\theta= - A_1 \sin \, \theta + A_2\cos \,\theta$. Indeed, the second relation in (\ref{Eq:AnsatzRadial}) simply corresponds to fixing the radial gauge. Plugging the ansatz \eqref{Eq:AnsatzRadial} into \eqref{Eq:EquationSecondOrdenCartesianCoordinates}, the following coupled nonlinear system of ordinary differential equations is obtained
\begin{equation}\label{Eq:betanfnEquations}
    \begin{split}
    \dfrac{d^2 f_n}{d r^2}+\frac{1}{r}\dfrac{d f_n}{dr }-\frac{n^2}{r^2}(1-\beta_n)^2 f_n+\dfrac{\lambda}{2}(1-f_n^2)f_n&=0,\\
    \dfrac{d^2\beta_n}{d r^2}-\frac{1}{r}\frac{d \beta_n}{d r}+(1-\beta_n) f_n^2 &=0.
    \end{split}
\end{equation}
 These equations must be complemented with the boundary conditions $f_n(0)=\beta_n(0)=0$, $f_n(\infty)=1$ and $\beta_n(\infty)=n$. In Figure \ref{Fig:Profiles} we show the profiles of the scalar and vector components of the 1-vortex for different values of the boundary constant $\lambda$.

\begin{figure}[ht!]
    \centering{
   \includegraphics[width=0.55\linewidth]{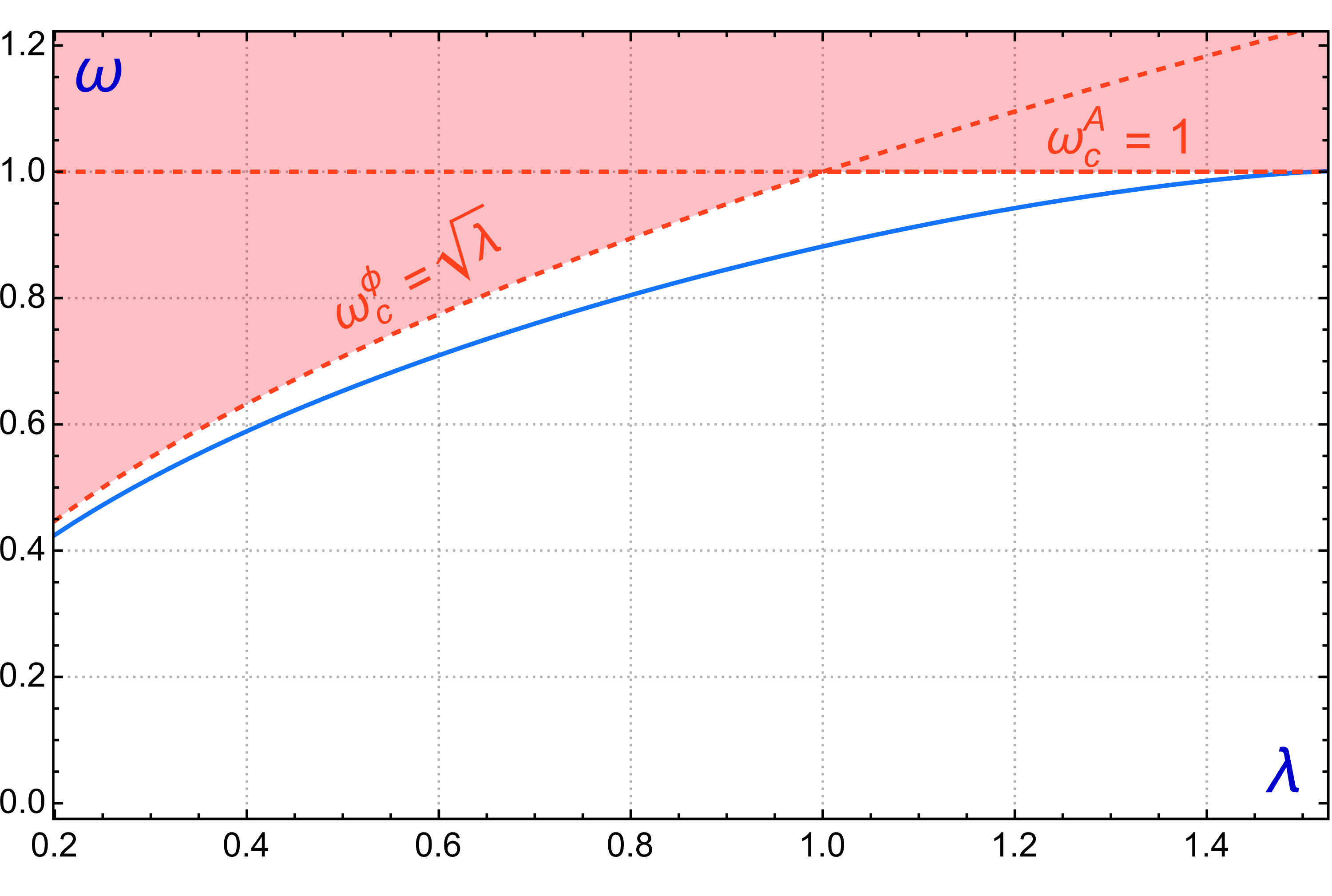}
    }
    \caption{Spectrum of the 1-vortex fluctuation operator $\mathcal{H}$, defined in \eqref{Eq:OperatorH}, as a function of the model parameter $\lambda$. The discrete eigenvalue corresponds to the blue curve, while the shaded red area represents the continuous spectrum.}
    \label{Fig:Spectrumn1}
\end{figure}

In order to identify the normal modes of the vortex we need to analyze the second-order small fluctuations operator ${\cal H}$. It has been demonstrated \cite{Izquierdo2016,AlonsoIzquierdo2016} that the lowest normal modes exhibit radial symmetry for vortex configurations with at least $n\leq 5$. Therefore, we will restrict our spectral analysis to fluctuations of the form $(\varphi(r),a_\theta(r))$\footnote{A more general ansatz for the internal mode structure along with the general eigenvalue problem can be found in \cite{Goodband1995}.} where $\varphi(r)$ and $a_\theta(r)$ denote respectively the fluctuations of the complex scalar field $\Phi$ and of the angular component of the vector field $A_\theta$. These perturbations have the same symmetry as the solutions derived from (\ref{Eq:betanfnEquations}) and satisfy the radial gauge condition, setting their radial component to zero, $a_r=0$. Therefore, the perturbed solution can be written as
\begin{equation}\label{Eq:Internalmodek0}
\Phi(r, \theta, t) = f_{n}(r)e^{i n \theta} + C_0\, \varphi(r)e^{i n \theta} \, e^{i \omega_n t} \hspace{0.4cm}, \hspace{0.4cm}  A_\theta(r, \theta, t) = \dfrac{n\,\beta_{n}(r)}{r} + C_0\,  a_\theta(r) \, e^{i \omega_n t} ,
\end{equation}
where $C_0$ is a small real number. Note that these fluctuations automatically satisfy the so-called background gauge condition \cite{Goodband1995,Izquierdo2016}. If we plugged (\ref{Eq:Internalmodek0}) into  \eqref{Eq:EquationSecondOrdenCartesianCoordinates}, the equations of motion at linear order in $C_0$ lead to the spectral problem 
\begin{equation}\label{Eq:SecondOrderOperator}
\mathcal{H}
\begin{pmatrix}
    \varphi(r) \\
    a_\theta (r) \\
\end{pmatrix}
= \omega_{n,j}^2
\begin{pmatrix}
   \varphi(r) \\
    a_\theta (r) \\
\end{pmatrix},
\end{equation}
where the subscript $n$ labels the vorticity of the configuration and $j$ labels the mode. The $n$-vortex fluctuation operator $\mathcal{H}$ reads
\begin{equation} \label{Eq:OperatorH}
    \mathcal{H}=
    \begin{pmatrix}
         -\dfrac{d^2}{dr^2}  - \dfrac{1}{r}\dfrac{d}{dr}  + \left(\dfrac{3}{2}\lambda f_{n}(r)^{2} -  \dfrac{\lambda}{2} + \dfrac{n^{2}}{r^{2}} - \dfrac{n^{2}\beta_{n}(r)}{r^{2}}\left( 2 - \beta_{n}(r) \right)\right) &  - \dfrac{2 n f_{n}(r)}{r}\left( 1 - \beta_{n}(r)\right)\\
         - \dfrac{2 n f_{n}(r)}{r}\left( 1 - \beta_{n}(r)\right) & - \dfrac{d^2}{dr^2}  - \dfrac{1}{r}\dfrac{d}{dr}  + \left(f_{n}(r)^2 + \dfrac{1}{r^2}\right)  
    \end{pmatrix}
    .
\end{equation}
The spectral problem (\ref{Eq:SecondOrderOperator}) couples the scalar and vector fluctuations. However, for large $r$ the fluctuation operator trivially decouples 
\begin{equation} \label{Eq:OperatorHInfinity}
    \mathcal{H}\big|_{\infty}=
    \begin{pmatrix}
         -\dfrac{d^2}{dr^2}  - \dfrac{1}{r}\dfrac{d}{dr}  + \lambda     &  0\\
        0 & - \dfrac{d^2}{dr^2}  - \dfrac{1}{r}\dfrac{d}{dr}  + \left(1 + \dfrac{1}{r^2}\right)  
    \end{pmatrix}
    .
\end{equation}
 The spectral problem given by the operator (\ref{Eq:OperatorHInfinity}) is analytically solvable. The (asymptotic) modes read
\begin{eqnarray}
    \varphi(r)&\xrightarrow[]{r\rightarrow\infty}&A_\varphi H_{0}^{(1)}(k_\phi r)+B_\varphi H_{0}^{(2)}(k_\phi r)\approx \sqrt{\dfrac{2}{\pi k_\phi r}}\left(A_\varphi e^{i(k_\phi r-\frac{\pi}{4})}+B_\varphi e^{-i(k_\phi r-\frac{\pi}{4})} \right), \label{Eq:AsymptoticPhi} \\
    a_\theta(r)&\xrightarrow[]{r\rightarrow\infty}&A_A H_{1}^{(1)}(k_A r)+B_A H_{1}^{(2)}(k_A r)\approx \sqrt{\dfrac{2}{\pi k_A r}}\left(A_A e^{i(k_A r-\frac{3\pi}{4})}+B_A  e^{-i(k_A r-\frac{3\pi}{4})} \right),\label{Eq:AsymptoticA}
\end{eqnarray}
where $k_\phi^2=\omega_{n,j}^2-\lambda$, $k_A^2=\omega_{n,j}^2-1$, and $H_n^{(1)}$, $H_n^{(2)}$ are the Hankel functions of first and second kind respectively \cite{Abramowitz1972,NIST2010}. From \eqref{Eq:OperatorHInfinity}, it is clear that the continuum spectrum for the vector fluctuations starts at the threshold value $\omega_c^A= 1$ while for the scalar component depends on the coupling constant and starts at $\omega_c^\phi=\sqrt{\lambda}$. Both of them coincide at critical coupling  ($\lambda=1$) where the self-dual vortices arise. As a consequence, bound states of (\ref{Eq:OperatorH}) must have eigenvalues $\omega_{n,j}^2<\lambda$ when $\lambda < 1$ and $\omega_{n,j}^2<1$ when $\lambda>1$. Due to the complexity of the spectral problem (\ref{Eq:SecondOrderOperator}) the discrete spectrum of (\ref{Eq:OperatorH}) must be obtained by employing numerical methods. The numerical scheme used in this paper is described in \ref{Sec:Appendix2}. In Figure \ref{Fig:Spectrumn1} we show the spectrum of the 1-vortex fluctuation operator (\ref{Eq:OperatorH}) as a function of the model parameter $\lambda$. We find that there is a single bound mode which ceases to exist at $\lambda\approx 1.5$.  

\begin{figure}[ht!]
    \centering
    
    \begin{subfigure}{0.489\textwidth}
        \includegraphics[width=\linewidth]{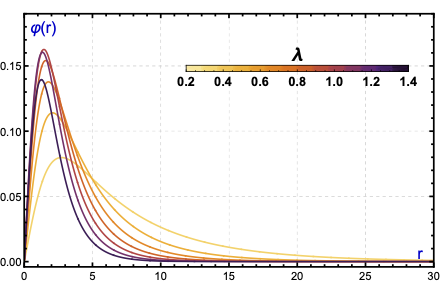}
    \end{subfigure}
    \hfill
    \begin{subfigure}{0.489\textwidth}
        \includegraphics[width=\linewidth]{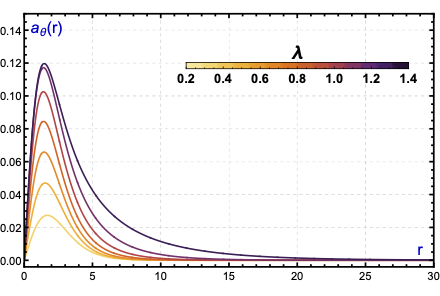}
    \end{subfigure}
  
    \caption{Scalar (left) and vector (right) component profiles of the 1-vortex bound mode for different values of the self-coupling constant $\lambda$.}
    \label{Fig:Profiles_BoundMode}
\end{figure}

Finally, in Figure \ref{Fig:Profiles_BoundMode} we show the profiles of the 1-vortex bound mode for different values of the self-coupling constant $\lambda$.

\section{Internal mode evolution: Analytical approach}\label{Sec:3}

In this section we will determine the decay law for a circularly symmetry vortex of charge $n$ with a single internal mode excited. Then, for concreteness, we will compare our analytical results with field theory for the 1-vortex.

The perturbative approach used to analyze the decay law of the wobbling amplitude for a kink was originally introduced in \cite{Manton1997}, and later extended to study kinks numerically in \cite{Blanco-Pillado:2020smt} as well as global vortices in \cite{BlancoPillado2021}. Here we employ similar techniques to derive the radiation emitted by an excited vortex. Then, by energy considerations, we will compute the decay law for the internal modes.

\subsection{Perturbative approach}

 We will employ the following circularly symmetric ansatz
\begin{eqnarray}
\Phi(r, \theta, t) &=& f_{n}(r)e^{i n \theta} + C(t)\varphi(r)e^{i n \theta} + \eta(r, t)e^{i n \theta}, \label{eq:phi_expansion} \\ 
A_{\theta}(r, t) &=& \dfrac{n\,\beta_{n}(r)}{r} + C(t) a_{\theta}(r) + \xi(r, t), \label{eq:a2_expansion}
\end{eqnarray}
where $\eta(r,t)$ and $\xi(r,t)$ represent the scalar and vector radiation field, $(\varphi(r), a_\theta(r))$  are the bound mode profiles and $C(t)$ is their time-dependent amplitude. A straightforward computation shows that $(\ref{eq:phi_expansion})$-$(\ref{eq:a2_expansion})$ still verify the temporal gauge and the background gauge conditions. 

To investigate the asymptotic radiation emitted due to the excitation of the internal mode, we insert the radially perturbed solution (\ref{eq:phi_expansion})-(\ref{eq:a2_expansion}) into the field equations (\ref{Eq:EquationSecondOrdenCartesianCoordinates}). The massive bound mode satisfies by definition the equations at first order in $C(t)$. However, the modes couple at higher order to radiation. Expanding at second order in $C(t)$ we derive the following equation of motion for the scalar field radiation
\begin{eqnarray}\label{eq:diff_eta_k0}
& & \Ddot{\eta}(r,t) - \eta''(r,t) - \dfrac{1}{r}\eta'(r,t)  + \left(\dfrac{3}{2}\lambda f_{n}(r)^{2} -  \dfrac{\lambda}{2} + \dfrac{n^{2}}{r^{2}} - \dfrac{n^{2}\beta_{n}(r)}{r^{2}}\left( 2 - \beta_{n}(r) \right)\right)\eta(r,t) - \nonumber \\
& &  - \dfrac{2 n f_{n}(r)}{r}\left( 1 - \beta_{n}(r)\right)\xi(r,t) = \frac{2n C(t)^2 \varphi(r)a_{\theta}(r)}{r}(1-\beta (r)) - C(t)^2 f_{n}(r)a_{\theta}(r)^2 -  \\
& & - \frac{3}{2}\lambda C(t)^2 f_{n}(r)\varphi(r)^2 - \left(\Ddot{C}(t) + \omega_s^2 C(t)\right)\varphi(r)~.\nonumber
\end{eqnarray}
For the gauge field radiation we get
\begin{eqnarray}\label{eq:diff_xi_k0}
& & \Ddot{\xi}(r,t) - \xi''(r,t) - \dfrac{1}{r}\xi'(r,t) + \left(f_{n}(r)^2 + \dfrac{1}{r^2}\right) \xi(r,t) - \dfrac{2n f_{n}(r)}{r}\left(1 - \beta_{n}(r) \right)\eta(r,t) = \nonumber\\ 
& & = \dfrac{n C(t)^2 \varphi^2(r)}{r}\left(1 - \beta_{n}(r)\right) - 2 C(t)^2 f_{n}(r)\varphi(r)a_{\theta}(r) - \left(\Ddot{C}(t) + \omega_s^2 C(t)\right) a_{\theta}(r)~,
\end{eqnarray}
where we have used the first order solution for $C(t)$
\begin{equation}\label{eq:first}
C(t) = C_0\cos(\omega_{n,j} t)~. 
\end{equation}
Using the fact that the discrete modes are orthogonal to the radiation modes we can now project (\ref{eq:diff_eta_k0}) and (\ref{eq:diff_xi_k0}) onto $\varphi(r)$ and $a_\theta(r)$ to obtain,

\begin{equation}\label{eq:C_02}
\Ddot{C}(t) + \omega_{n,j}^2 C(t) + C(t)^2\gamma = 0~,
\end{equation}
where
\begin{equation} \label{eq:norma}
\gamma = \dfrac{2 \pi}{\mathcal{N}}\int_{0}^{\infty}\bigg( \dfrac{3}{2}\lambda f_{n}(r)\varphi(r)^3 + 3f_{n}(r)\varphi(r)a_{\theta}(r)^2 - \dfrac{3 n \varphi(r)^2 a_{\theta}(r)}{r}(1 - \beta_{n}(r)) \bigg)\,r dr~,
\end{equation}
and $\mathcal{N}$ in (\ref{eq:norma}) is the normalization factor of the shape mode
\begin{equation}\label{eq:factor_N}
\mathcal{N} = 2\pi \int_{0}^{\infty}\left(\varphi(r)^2 + a_{\theta}(r)^2\right) r\, dr \,  ~.
\end{equation} 

Furthermore, we will assume the following ansatz for the radiation
modes
\begin{eqnarray}\label{eq:rad_1}
\eta(r, t) &=& \eta_{r}(r)\,e^{i \omega t}~,\\ \label{eq:rad_2}
\xi(r,t) &=& \xi_{r}(r)\,e^{i \omega t}~,
\end{eqnarray}
which, together with first order solution (\ref{eq:first}) leads to the condition $\omega = 2\omega_{n,j}$, i.e. the radiation frequency in twice that of the discrete mode. After substituting (\ref{eq:rad_1}), (\ref{eq:rad_2}) and (\ref{eq:C_02}) into (\ref{eq:diff_eta_k0}) and (\ref{eq:diff_xi_k0}) we get the following equations

\begin{eqnarray}\label{eq:diff_eta_2}
& & - \eta_r''(r) - \dfrac{1}{r}\eta_r'(r)  + \left(\dfrac{3}{2}\lambda f_{n}(r)^{2} -  \dfrac{\lambda}{2} + \dfrac{n^{2}}{r^{2}} - 4\omega_{n,j}^2 - \dfrac{n^{2}\beta_{n}(r)}{r^{2}}\left( 2 - \beta_{n}(r) \right)\right)\eta_{r}(r) - \dfrac{2 n f_{n}(r)}{r}\left( 1 - \beta_{n}(r)\right)\xi_r(r) = \nonumber \\
& & = \frac{n C_0^2 \varphi(r)a_{\theta}(r)}{r}(1-\beta_{n}(r)) - \dfrac{1}{2}C_0^2 f_{n}(r)a_{\theta}(r)^2 - \frac{3}{4}\lambda C_0^2 f_{n}(r)\varphi(r)^2 + \dfrac{1}{2}C_0^2\gamma \varphi(r) \equiv C_0^2 F_{\phi}(r)~,
\end{eqnarray}
and 
\begin{eqnarray}\label{eq:diff_xi_2}
& & - \xi_r''(r) - \dfrac{1}{r}\xi_r'(r) + \left(f_{n}(r)^2 + \dfrac{1}{r^2} - 4\omega_{n,j}^2\right) \xi_r(r) - \dfrac{2n f_{n}(r)}{r}\left(1 - \beta_{n}(r) \right)\eta_{r}(r) = \nonumber\\ 
& & = \dfrac{n C_0^2 \varphi^2(r)}{2r}\left(1 - \beta_{n}(r)\right) - C_0^2 f_{n}(r)\varphi(r)a_{\theta}(r) + \dfrac{1}{2}C_0^2\gamma a_{\theta}(r) \equiv C_0^2 F_{A}(r)~.
\end{eqnarray}

The equations (\ref{eq:diff_eta_2})-(\ref{eq:diff_xi_2}) constitute a coupled system of non-homogeneous linear ordinary differential equations. It should be noted, however, that the cross-terms exponentially vanish for large $r$. In order to solve the system we may use an iterative approach  similar to Bohr's approximation. The procedure works as follows: First, we apply the method of variation of parameters as if the coupling terms were part of the inhomogeneous terms. Then, a particular solution is given by
\begin{eqnarray}
\eta_{r}^{(m)}(r) &=& - C_0^2\, z_{2\phi}(r) \int_{0}^{r} \left(F_{\phi}(r') - \dfrac{2 n f_{n}(r')}{r'}\left( 1 - \beta_{n}(r')\right)\xi_r^{(m - 1)}(r') \right) \dfrac{z_{1\phi}(r')}{W_{\phi}(r')}dr'\nonumber \\ 
& & - C_0^2\, z_{1\phi}(r) \int_{r}^{\infty} \left(F_{\phi}(r') - \dfrac{2 n f_{n}(r')}{r'}\left( 1 - \beta_{n}(r')\right)\xi_r^{(m - 1)}(r') \right) \dfrac{z_{2\phi}(r')}{W_{\phi}(r')}dr',\label{eq:VarPar_eta_k0}\\
\xi_r^{(m)}(r) &=& - C_0^2\, z_{2 A}(r) \int_{0}^{r} \left(F_{A}(r') - \dfrac{2 n f_{n}(r')}{r'}\left( 1 - \beta_{n}(r')\right)\eta_{r}^{(m - 1)}(r') \right) \dfrac{z_{1 A}(r')}{W_{A}(r')}dr'\nonumber\\ 
& & - C_0^2\, z_{1 A}(r) \int_{r}^{\infty} \left(F_{A}(r') - \dfrac{2 n f_{n}(r)}{r}\left( 1 - \beta_{n}(r)\right)\eta_{r}^{(m - 1)}(r') \right) \dfrac{z_{2 A}(r')}{W_{A}(r')}dr', \label{eq:VarPar_xi_k0}
\end{eqnarray}
where  $z_{j\phi}$ and $z_{j A}$ ($j=1,2$) respectively denote the two linearly independent homogeneous solutions for the scalar and vector components $\eta$ and $\xi$ of (\ref{eq:diff_eta_2})-(\ref{eq:diff_xi_2}), and $F_{\phi}$ and $F_{A}$ account for their non-homogeneous terms as indicated in (\ref{eq:diff_eta_2})-(\ref{eq:diff_xi_2}). Besides, $W_{\phi}$ and $W_{A}$ are respectively the Wronskians associated to the homogeneous solutions $z_{j\phi}$ and $z_{j A}$, $j=1,2$. The index $m$ in (\ref{eq:VarPar_eta_k0})-(\ref{eq:VarPar_xi_k0}) indicates the iteration step, with $\eta_{r}^{(0)}(r)=0$ and $\xi_r^{(0)}(r)=0$. Although the solutions of the homogeneous system associated to (\ref{eq:diff_eta_2})-(\ref{eq:diff_xi_2}) cannot be analytically identified, it is possible to find their asymptotic behaviors (the equations decouple in this limit as shown in Section \ref{Sec:2}). We have that
\begin{eqnarray}
& & z_{\phi}(r) \xrightarrow{r \longrightarrow \infty} c_1 J_0(q_{\phi} \, r) + c_2 Y_0(q_{\phi}\, r ),\\  
& & z_{A}(r) \xrightarrow{r \longrightarrow \infty} d_1 J_1(q_{A}\, r ) + d_2 Y_1(q_{A}\, r )~,
\end{eqnarray}
where $J_n$ and $Y_n$ are the Bessel $J$ and Bessel $Y$ functions, respectively, and 
\begin{equation}
q_{\phi} = \sqrt{4 \,\omega_{n,j}^2 - \lambda}, \quad q_{A} = \sqrt{4 \, \omega_{n,j}^2 - 1}~.
\end{equation}
We choose the asymptotic radiation to be outgoing waves in the radial direction, then 
\begin{eqnarray}
z_{1\phi}(r) & \xrightarrow{r \longrightarrow \infty} & \Tilde{c}_1\,J_0(q_{\phi} \,r )~, \hspace{0.3cm} z_{2\phi}(r)  \xrightarrow{r \longrightarrow \infty} H^{(2)}_{0}(q_{\phi} \, r)\label{eq:asym_eta}~, \\
z_{1 A}(r) & \xrightarrow{r \longrightarrow \infty} & \Tilde{d}_2\,Y_1(r q_{A})~, \hspace{0.3cm} z_{2 A}(r) \xrightarrow{r \longrightarrow \infty} H^{(2)}_{1}(rq_{A})\label{eq:asym_xi}~,
\end{eqnarray}
where, once again, $H_n^{(2)}$ denote the Hankel function of second kind. The asymptotic expansion of (\ref{eq:VarPar_eta_k0})-(\ref{eq:VarPar_xi_k0}) reduces to 
\begin{eqnarray}
\eta_{r}^{(m)}(r) &=& - C_0^2\,\sqrt{\dfrac{2}{\pi r q_{\phi}}}  \cdot \, I^{(m)}_{\phi} \cdot  \, e^{-i r q_{\phi}+\frac{i \pi }{4}} , \label{eq:int_eta}\\ 
\xi_r^{(m)}(r) &=& - C_0^2\,\sqrt{\dfrac{2}{\pi r q_{A}}}  \cdot \, I^{(m)}_{A} \cdot  \, e^{-i r q_{A}+\frac{3 i \pi }{4}} \label{eq:int_xi},
\end{eqnarray}
where the factors $I^{(m)}_{\phi}$ and $I^{(m)}_{A}$ are defined by the following integrals
\begin{eqnarray}
I^{(m)}_{\phi} &=& \int_{0}^{\infty} \left(F_{\phi}(r') - \dfrac{2 n f_{n}(r')}{r'}\left( 1 - \beta_{n}(r')\right)\xi_r^{(m - 1)}(r') \right) \dfrac{z_{1\phi}(r')}{W_{\phi}(r')}dr', \label{integral1}\\
I^{(m)}_{A} &=& \int_{0}^{\infty} \left(F_{A}(r') - \dfrac{2 n f_{n}(r')}{r'}\left( 1 - \beta_{n}(r')\right)\eta_{r}^{(m - 1)}(r') \right) \dfrac{z_{1 A}(r')}{W_{A}(r')}dr'. \label{integral2}
\end{eqnarray}

To compute the integrals (\ref{integral1}) and (\ref{integral2}), it is necessary to identify numercally the homogeneous solutions $z_{1\phi}$ and $z_{1 A}$ with the boundary conditions $z_{1\phi}(0) = z_{1 A}(0) = 0$ and those given by the asymptotic behavior (\ref{eq:asym_eta}) and (\ref{eq:asym_xi}) (see Reference \cite{BlancoPillado2021}). After some straightforward algebraic manipulations we finally get
\begin{eqnarray}
\eta^{(m)}(r,t) &\approx& C_0^2\, {\rm Re} \, \left[\dfrac{C^{(m)}_{\phi}}{\sqrt{r}}e^{i2\omega_{n,j} t - i r q_{\phi}+\frac{i \pi }{4}} \right] = C_0^2\dfrac{C^{(m)}_{\phi}}{\sqrt{r}}\,\cos \Big(2\omega_{n,j} t - r q_{\phi} + \dfrac{\pi}{4} + \zeta_{\phi} \Big) \label{eq:rad_eta}~, \\
\xi^{(m)}(r,t) &\approx& C_0^2\, {\rm Re} \, \left[\dfrac{C^{(m)}_{A}}{\sqrt{r}}e^{i2\omega_{n,j} t - i r q_{A}+\frac{3i \pi }{4}} \right] = C_0^2\dfrac{C^{(m)}_{A}}{\sqrt{r}}\,\cos \Big (2\omega_{n,j} t - r q_{A} + \dfrac{3\pi}{4} + \zeta_A \Big) \label{eq:rad_xi}~,
\end{eqnarray}
where the normalized radiation amplitudes $C^{(m)}_{\phi}$ and $C^{(m)}_{A}$ are given by
\begin{equation} \label{eq:C_amplitudes} 
C^{(m)}_{\phi} = \sqrt{\dfrac{2}{\pi q_{\phi}}} |I^{(m)}_{\phi}| \hspace{0.5cm} \mbox{and} \hspace{0.5cm} C^{(m)}_{A} = \sqrt{\dfrac{2}{\pi  q_{A}}}|I^{(m)}_{A}| ~.
\end{equation}
The quantities $\zeta_{\phi}$ and $\zeta_A$ in (\ref{eq:rad_eta}) and (\ref{eq:rad_xi}) are the phases of the scalar and vector radiation irrelevant to our purposes. From equations $(\ref{eq:rad_eta})$-$(\ref{eq:rad_xi})$, it can be observed that the radiation amplitudes are proportional to the square of the shape mode amplitude $C_0^2$, consistent with our initial assumption. The asymptotic radiation profiles formally resemble the analogous expression for the global vortex \cite{BlancoPillado2021}. However, unlike in that case, there are two radiation channels (scalar and vector) which are sensitive to the self-coupling $\lambda$. We will discuss this is detail in the following sections.

\subsection{Internal mode decay law}

The asymptotic form of the radiation (\ref{eq:rad_eta})-(\ref{eq:rad_xi}), allows us to determine the decay law of the internal mode excitations. This can be done by comparing the average energy flux carried away by the radiation with the rate of change of the energy of the excited vortex. To begin, we compute the energy flux in the radial direction, represented by the $T_{0r}$ component of the energy momentum tensor
\begin{equation}
T_{0r} = \dot{\eta}(r,t)\partial_{r}\eta(r,t) + \dot{\xi}(r,t)\partial_{r}\xi(r,t) + \dot{\xi}(r,t)\dfrac{\xi(r,t)}{r}~.
\end{equation}
The average energy flux over one period is given by
\begin{equation}
\langle T_{0r} \rangle = - \dfrac{2 C_0^4 \omega_{n,j}}{\pi r}\left(|I^{(m)}_{\phi}|^2 + |I^{(m)}_{A}|^2\right)~.
\end{equation}
Thus, the power radiated to infinity is
\begin{equation}
\int_{0}^{2\pi}\, \langle T_{0r} \rangle r\,d\theta = - 4 C_0^4 \omega_{n,j}\left(|I^{(m)}_{\phi}|^2 + |I^{(m)}_{A}|^2\right)~.
\end{equation}
On the other hand, the energy associated to the vibrating vortex (excited only by the internal mode) is given by
\begin{equation}
E = M_V + \dfrac{1}{2}\mathcal{N}\omega_{n,j}^2 C_0^2~,
\end{equation}
where $M_V$ denotes the rest mass of the gauged vortex and $\mathcal{N}$ is the normalization factor previously defined in $(\ref{eq:factor_N})$. Therefore, the energy associated to the vibrational mode depends on the squared amplitude $C_0^2$, so while the vibrating vortex emits radiation, the amplitude must decrease over time to maintain the energy balance. This implies that
\begin{equation}  \label{edoCdet}
\dfrac{\mathcal{N}}{2}\omega_{n,j}\dfrac{d C_0(t)^2}{dt} = - 4C_0(t)^4 \left(|I^{(m)}_{\phi}|^2 + |I^{(m)}_{A}|^2\right).
\end{equation}
The solution of (\ref{edoCdet}) provides the decay law for the internal mode amplitude 
\begin{equation}\label{eq:decay_law}
C_0(t) = \dfrac{1}{\sqrt{C_0^{-2}(0) + \Gamma_{n,j}^{(m)}  t}}~,
\end{equation}
where $C_0(0)=C_0$ is the initial shape mode amplitude (at $t=0$) and 
\begin{equation}\label{eq:Gamma}
\Gamma_{n,j}^{(m)} = \dfrac{8 \left(|I^{(m)}_{\phi}|^2 + |I^{(m)}_{A}|^2\right)}{\mathcal{N}\omega_{n,j}}~,
\end{equation}
is the decay rate. The relation $(\ref{eq:decay_law})$ is formally identical to the decay law for the amplitude of a kink \cite{Manton1997} or for a global vortex \cite{BlancoPillado2021}, but here both the scalar and vector radiation field contribute to the final expression.

\begin{figure}[ht!]
    \centering{
   \includegraphics[width=0.55\linewidth]{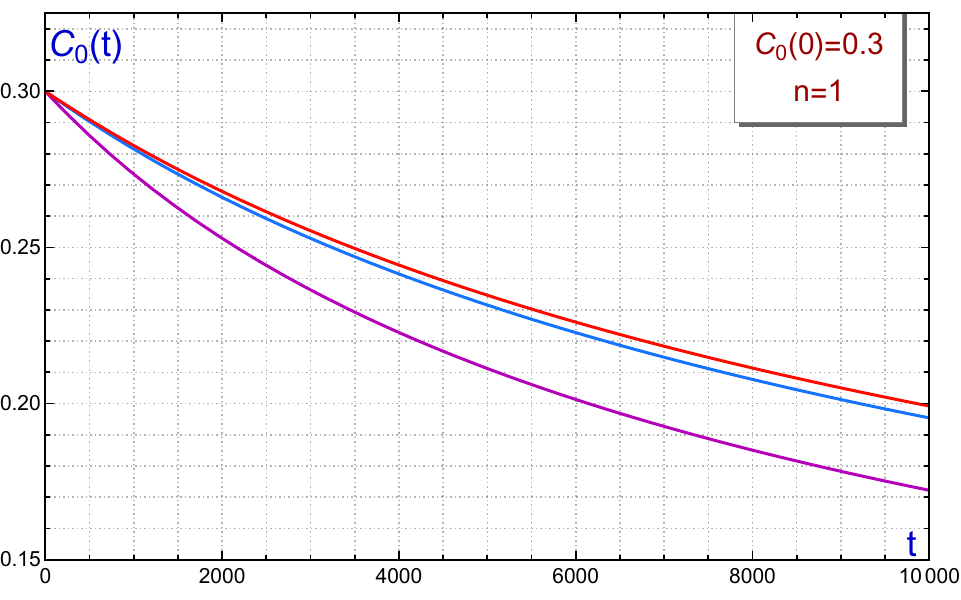}}
    \caption{Graphical representation for the theoretical decay of $C_0$ described by equation \eqref{eq:decay_law}  for $\lambda = 0.7$ (blue), $\lambda = 1$ (purple), and $\lambda = 1.4$ (red) . }\label{Fig:AmpVsTimeTeor}
\end{figure}

\begin{figure}[ht!]
    \centering{
   \includegraphics[width=0.55\linewidth]{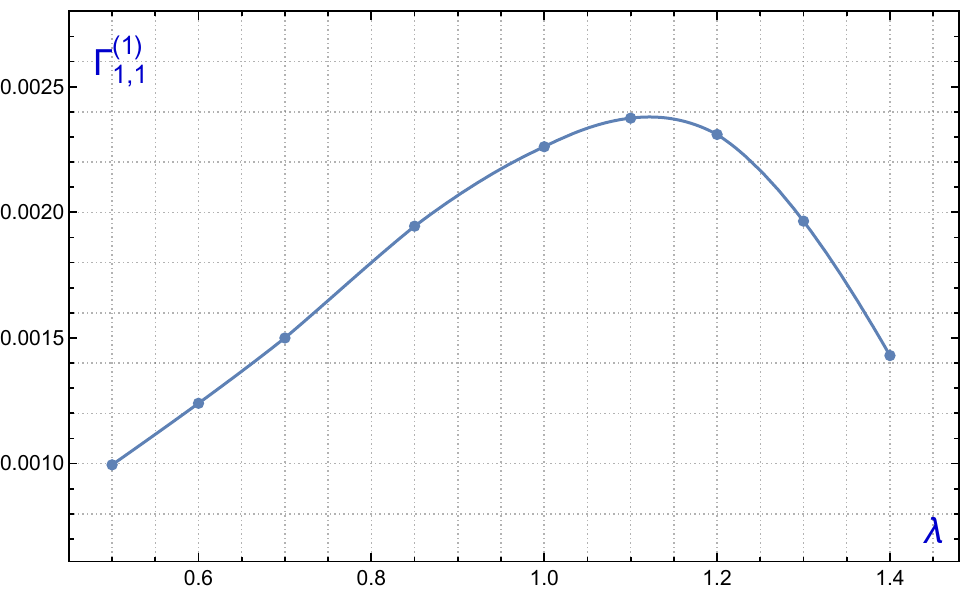}
    }
    \caption{Graphical representation of the decay rate $\Gamma_{1,1}^{(1)}$ as a function of the coupling constant $\lambda$ for 1-vortices. The points depict the analytical results derived from equation $(\ref{eq:Gamma})$, while the solid line illustrates a numerical interpolation.}
    \label{Fig:Gamma_Lambda}
\end{figure}

Figure \ref{Fig:AmpVsTimeTeor} displays the evolution of the shape mode amplitude $C_0(t)$ for 1-vortices for the values of $\lambda = 0.7$, $\lambda = 1$ and $\lambda = 1.4$. To obtain a comprehensive pattern of the decay of the shape mode amplitude (\ref{eq:decay_law}) a representation illustrating the dependence of the decay rate $\Gamma$ on $\lambda$ can be found in Figure \ref{Fig:Gamma_Lambda}. 
We have taken $\lambda > 0.5$
since reducing $\lambda$ further would increase the size of the vortex core making necessary to use a substantially larger simulation box.
For this range of values the decay rate reaches a maximum for $\lambda$ close to $1$, that is, close to the critical value. This is related to the finite size of the source. Around the maximum of the decay rate the vortex size and the masses of the radiated particles are of the same order. Far from this point the difference in scales of the size of the object and the masses of the radiated particles should lead to radiation suppression \cite{BlancoPillado2021}.

\section{Internal mode evolution: Numerical approach}  \label{Sec:4}
 
In the preceding section, we have employed perturbation theory to derive the asymptotic behavior of radiation in both the scalar and vector channels, enabling us to obtain an analytical expression that determines the decay of the shape mode amplitude. In this section, we carry out a similar investigation using numerical methods. The outcomes obtained through this approach will be compared with those from the preceding section, allowing us to validate the accuracy of the previously derived expressions.

\subsection{Numerical setup and radiation power spectrum analysis}\label{Sec:4.2}

As mentioned in Section \ref{Sec:2}, we will restrict our analysis to circularly symmetric configurations. Therefore, the system of differential equations to be numerically solved is as follows:
\begin{eqnarray}
\dfrac{\partial^2 F_n}{\partial t^2} - \dfrac{\partial^2 F_n}{\partial r^2} - \frac{1}{r}\dfrac{\partial F_n}{\partial r } + \frac{n^2}{r^2}(1-B_{n})^2 F_n - \dfrac{\lambda}{2}(1-F_n^2)F_n&=0~, \label{eq:TemporalEvolution1}\\
\dfrac{\partial^2 B_{n}}{\partial t^2} - \dfrac{\partial^2 B_{n}}{\partial r^2} + \frac{1}{r}\frac{\partial B_{n}}{\partial r} - (1-B_{n}) F_n^2 &=0~, \label{eq:TemporalEvolution2}
\end{eqnarray}
where $F_n(r,t)$ denotes the vortex profile of vorticity $n$ and $B_n(r,t) = \frac{r}{n} A_{\theta}(r,t)$ denotes the angular component of the gauge field. To solve this system, a second-order finite difference scheme in both space and time has been implemented with $\Delta r = 0.01$ and $\Delta t = 0.001$. The simulations were performed up to $t=10000$ over the radial interval $[0,L]$, where $L = 70$. We note that using larger boxes did not yield substantial differences in our simulation. To prevent radiation reflecting from the boundary at $r = L$, Mur absorbing boundary conditions have been introduced \cite{Mur}. Additionally, we introduced damping terms $-\epsilon(r) \frac{d F_n}{dt}$ and $-\epsilon(r) \frac{d B_n}{dt}$ respectively in (\ref{eq:TemporalEvolution1}) and (\ref{eq:TemporalEvolution2}) where

\begin{equation}
\epsilon(r) = \begin{cases}
0, & \text{if } 0 \leq r < r_{\text{cut}}~, \\
\left[\dfrac{r - r_{\text{cut}}}{20}\right]^4, & \text{if } r_{\text{cut}} < r < L~.
\end{cases}
\end{equation}
 Typical values have been chosen around $r_{cut} = 5L/6$, but we have performed several tests at different values of $r_{cut}$ to confirm the numerical stability of our numerical setup. A large number of simulations have been carried out to examine the evolution of a vibrating 1-vortex. Moreover, the eigenfunctions are assumed normalized with respect to the norm of $L^2(\mathbb{R}^2)\oplus\mathbb{R}^2$. In all numerical simulations we have focussed on the $n=1$ vortex. The evolution of $n-$vortices with $n>1$ will be explored in future studies, as they introduce some subtleties that require special analysis.

\begin{figure}[ht!]
    \centering
    \begin{subfigure}{0.489\textwidth}
        \includegraphics[width=\linewidth]{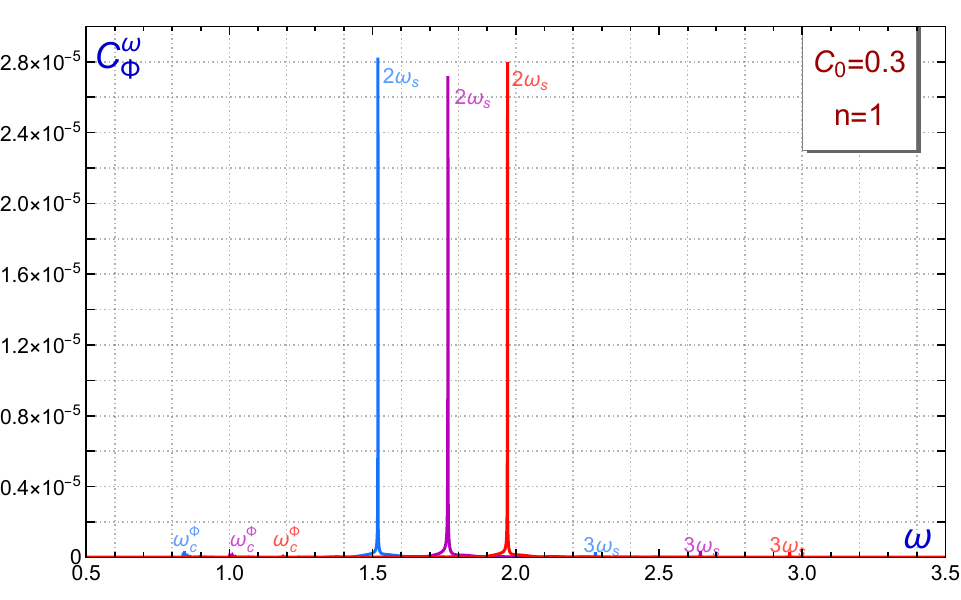}
        \caption{Scalar channel.}
        \label{Fig:Power_Phi}
    \end{subfigure}
    \hfill
    \begin{subfigure}{0.489\textwidth}
        \includegraphics[width=\linewidth]{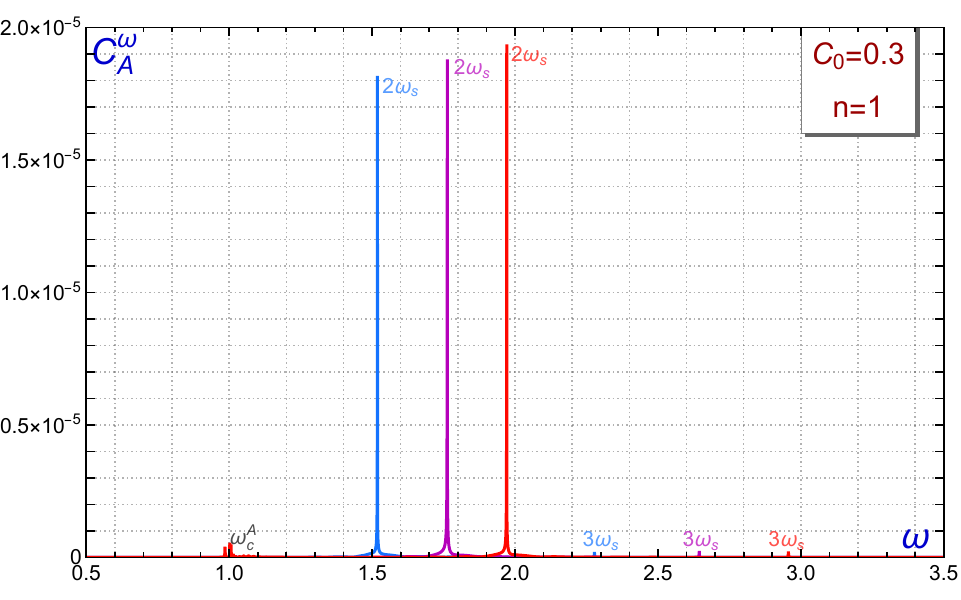}
        \caption{Vector channel.}
        \label{Fig:Power_A}
    \end{subfigure}
  
    \caption{Power spectra of the scalar (left) and vector (right) radiation fields for $\lambda = 0.7$ (blue), $\lambda = 1$ (purple), and $\lambda = 1.4$ (red). The peak labels indicate the corresponding frequencies in each case, where $\omega_c^{\phi} = \sqrt{\lambda}$ denotes the scalar threshold frequency and $\omega_c^{A} = 1$ represents the gauge mass threshold frequency. The initial amplitude for the massive bound mode is set to $C_0 = 0.3$. }
    \label{Fig:Power_Spectrum}
\end{figure}

The first analysis of the data extracted from these simulations is aimed at studying the frequencies of the radiation emitted by the 1-vortex at a point far away from the vortex core both in the scalar and vector channels. The results found for the model parameters $\lambda = 0.7, \lambda = 1.0$, and $\lambda = 1.4$ are illustrated in Figure \ref{Fig:Power_Spectrum}. The power spectrum of the radiation emitted by the gauge vortex in the scalar channel has been depicted in Figure \ref{Fig:Power_Phi} . As anticipated, the predominant peak occurs at $\omega(\lambda) = 2\omega_s(\lambda)$. Additionally, other peaks arise although comparatively suppressed. For example, small peaks around $3\omega_s(\lambda)$ are observed, due to the coupling between the internal and radiation modes at higher orders. 

The presence of these other peaks is a higher order effect and therefore is not included in our perturbative approach based on the dominant second order expansion. Nevertheless, they might be captured if we expanded up to the third order. Lastly, there are other minor peaks denoted by $\omega_c^{\phi} = \sqrt{\lambda}$ caused by random numerical noise. Analogously, Figure \ref{Fig:Power_A} shows the power spectrum of the radiation emitted in the vector channel, which exhibits a similar behaviour.

\begin{figure}[ht!]
    \centering
    
    \begin{subfigure}{0.489\textwidth}
        \includegraphics[width=\linewidth]{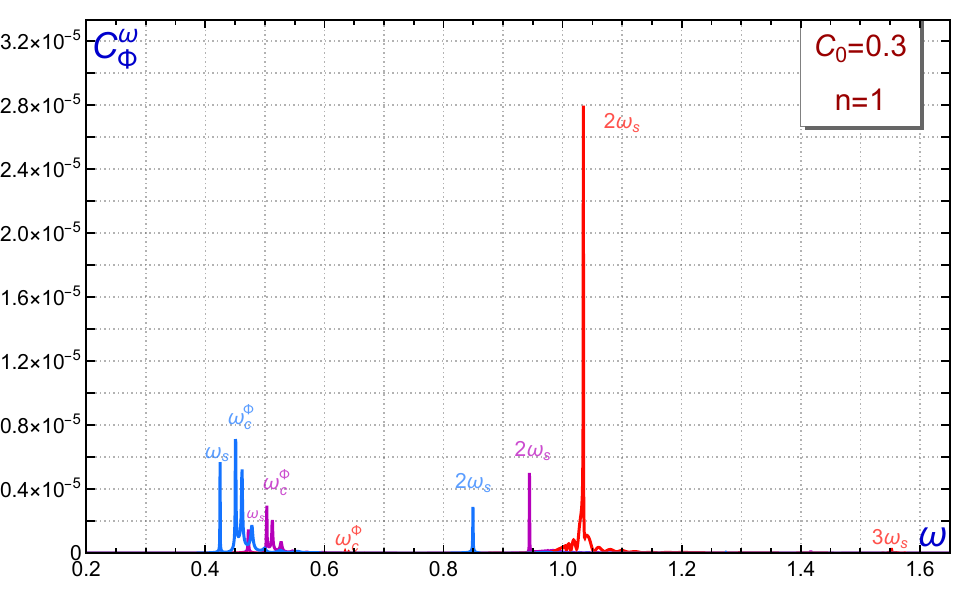}
        \caption{Scalar channel.}
        \label{Fig:Power_Phi_low}
    \end{subfigure}
    \hfill
    \begin{subfigure}{0.489\textwidth}
        \includegraphics[width=\linewidth]{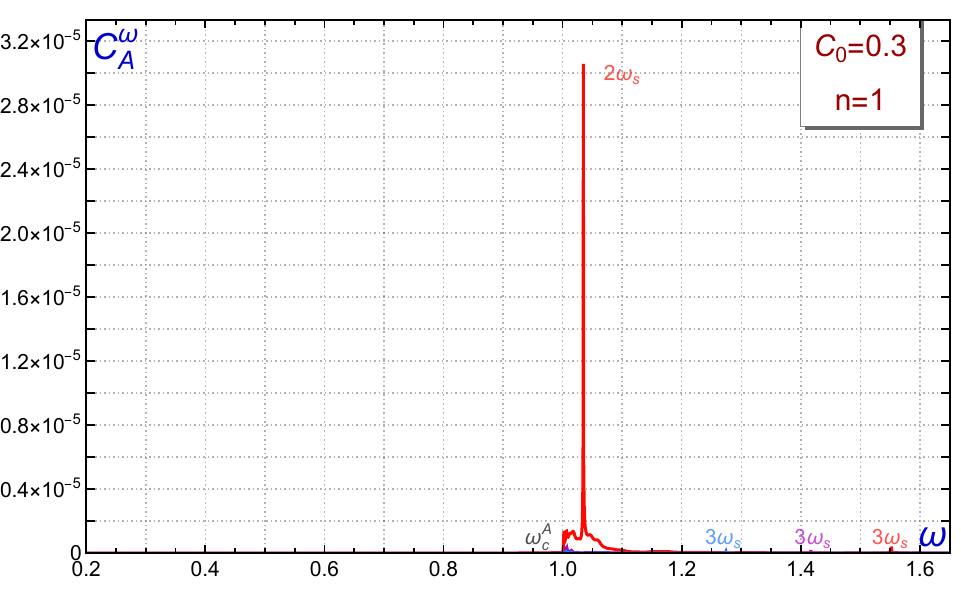}
        \caption{Vector channel.}
        \label{Fig:Power_A_low}
    \end{subfigure}
  
    \caption{Power spectra of the scalar (left) and vector (right) radiation fields for $\lambda = 0.2$ (blue), $\lambda = 0.24$ (purple), and $\lambda = 0.4$ (red). The peak labels indicate the corresponding frequencies in each case, where $\omega_c^{\phi} = \sqrt{\lambda}$ denotes the scalar threshold frequency and $\omega_c^{A} = 1$ represents the gauge mass threshold frequency. The initial amplitude for the massive bound mode is set to $C_0 = 0.3$.}
    \label{Fig:Power_Spectrum_2}
\end{figure}

The scheme depicted by the power spectrum in Figure \ref{Fig:Power_Spectrum} is valid in the regime where $\lambda > 0.282$ and $\lambda < 1.5$, in which there exists only one discrete mode and its first harmonic belongs to the doubly degenerate continuous spectrum. A qualitative change occurs for $\lambda \lesssim 0.282$. In this case, the first harmonic $\omega(\lambda) = 2 \omega_s(\lambda)$ is higher than the Higgs mass but lower than the gauge mass (see Figure \ref{Fig:Spectrumn1}). Thus, it is expected that only the scalar channel is available to emit radiation. In Figure \ref{Fig:Power_Spectrum_2}, we display the power spectrum for $\lambda = 0.2$, $\lambda = 0.25$, and $\lambda = 0.4$. Note that the cases $\lambda = 0.2$ and $\lambda = 0.25$ belong to the regime where only the scalar channel is available at the lowest order while for the case $\lambda = 0.4$ both channels can radiate. As mentioned in Section \ref{Sec:2} we anticipate a different behavior in the radiation emission for these two scenarios. As expected, in Figure \ref{Fig:Power_A_low} the peak at twice the frequency of the corresponding shape mode in the vector channel for $\lambda = 0.2$ and $\lambda = 0.25$ is highly suppressed. Note that in Figure \ref{Fig:Power_Phi_low}, the presence of shape mode oscillations for the cases $\lambda = 0.2$ and $\lambda = 0.25$ are noticeable at large distances (where the power spectrum analysis is considered). This is because the shape mode for these cases has a significant width, and the power spectrum in the simulations captures these frequencies. However, it is interesting to describe in detail the situation arising in this case. It should be noted that since the eigenvalue of the discrete mode is very close to the threshold value of the continuous spectrum associated to the scalar component and comparatively far from that of the vector component, the scalar component of the shape mode will be highly delocalized and will have a dominant behavior over the vector component. This explains the fact that only Figure \ref{Fig:Power_Phi_low} shows frequencies associated to the discrete mode when $\lambda < 0.282$.

In summary, the numerical simulations described in this Section establish that the dominant frequency for the radiation emitted when a 1-vortex vibrates via its shape mode is precisely twice the natural frequency of the shape mode.

\subsection{Decay rate of the internal mode and radiation emission}

In this section, we will analyze the numerical decay of the shape mode amplitude and compare it with the analytical expression (\ref{eq:decay_law}) obtained in Section \ref{Sec:3}. 
The numerical amplitude of the shape mode at each time $t$ is computed in our simulations by projecting the difference between the evolving and the static 1-vortex onto the theoretical shape mode, as follows:
\begin{equation}
C(t) \approx 2 \pi \int_{0}^{\infty}\left[\left(F_n(r,t) - f_n(r)\right)\varphi(r) + \frac{n}{r}\left(B_{n}(r,t) - \,\beta_n(r)\right)a_{\theta}(r)\right]r\, dr,
\end{equation}
which is justified taking into account the expressions (\ref{eq:phi_expansion}) and (\ref{eq:a2_expansion}).

\begin{figure}[ht!]
    \centering
    
    \begin{subfigure}{1\textwidth}
        \includegraphics[trim={3cm 0 3cm 0},clip,width=\linewidth]{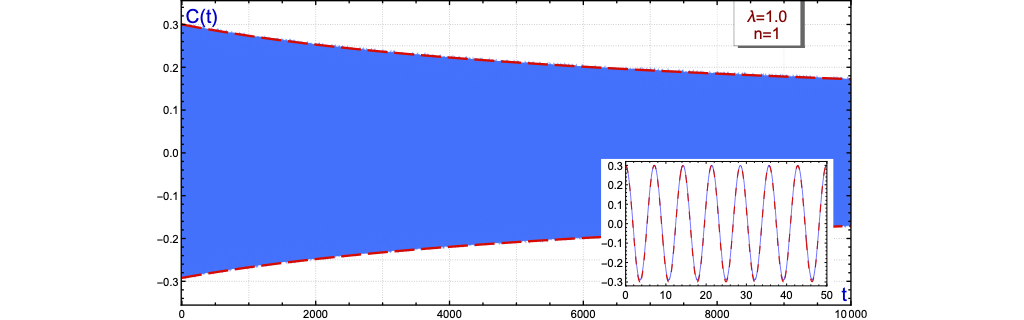}
        \caption{$\lambda = 1$.}
    \end{subfigure} \hspace{0.2cm}
    \begin{subfigure}{0.41\textwidth}
        \includegraphics[width=\linewidth]{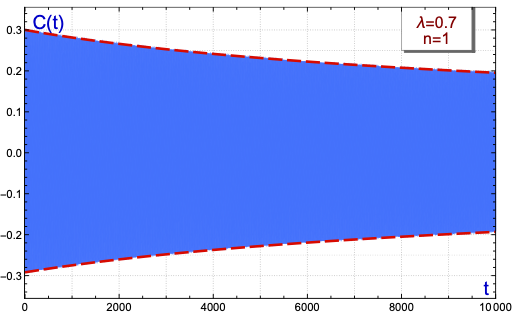}
        \caption{$\lambda = 0.7$.}
    \end{subfigure}\hspace{0.2cm}
    \begin{subfigure}{0.41\textwidth}
        \includegraphics[width=\linewidth]{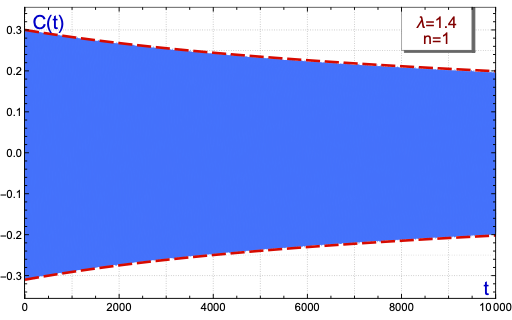}
        \caption{$\lambda = 1.4$.}
    \end{subfigure}
    
    \caption{Evolution of the numerical shape mode amplitude (blue solid curve) and the analytical decay law $(\ref{eq:decay_law})$ (red dashed curve) for the coupling constants $\lambda=0.7$, $\lambda=1.0$ and $\lambda=1.4$. The shape mode amplitude decays due to the coupling with scattering modes. All the simulations have been performed for an initial shape mode amplitude $C_0 = 0.3$. The inset  shows the first oscillations of the internal mode amplitude.}
    \label{Fig:Decay_Law}
\end{figure}

The numerical amplitude of the shape mode as a function of $t$ exhibits a large number of oscillations during our simulations. These oscillations are depicted by the blue curves in Figure \ref{Fig:Decay_Law} for the values $\lambda=0.7$, $\lambda=1$, and $\lambda=1.4$. However, due to the ratio between their oscillation periods and the total simulation time, they appear as a continuous blue area. On the other hand, the analytical response of the shape mode amplitudes (\ref{eq:decay_law}) are represented by the dashed red curves for the same cases in Figure \ref{Fig:Decay_Law}. This figure shows a close match between the envelope of the numerical oscillations of the shape mode and the analytical amplitude.

\begin{figure}[ht!]
    \centering
    
    \begin{subfigure}{0.328\textwidth}
        \includegraphics[width=\linewidth]{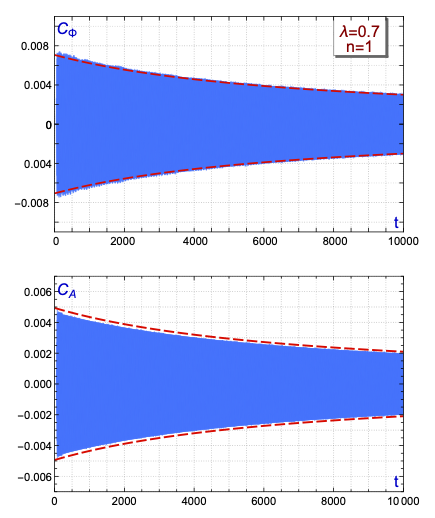}
        \caption{$\lambda = 0.7$.}
    \end{subfigure}
    \hfill
    \begin{subfigure}{0.328\textwidth}
        \includegraphics[width=\linewidth]{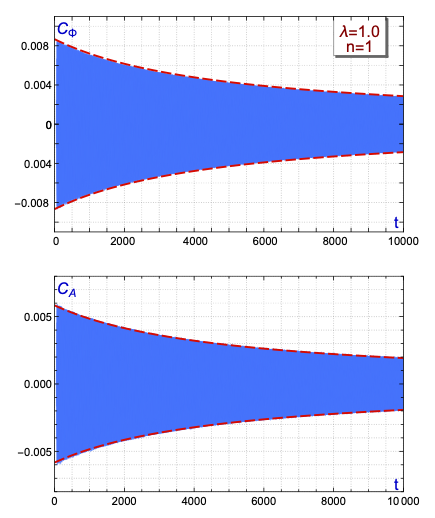}
        \caption{$\lambda = 1.0$.}
    \end{subfigure}
    \hfill
    \begin{subfigure}{0.328\textwidth}
        \includegraphics[width=\linewidth]{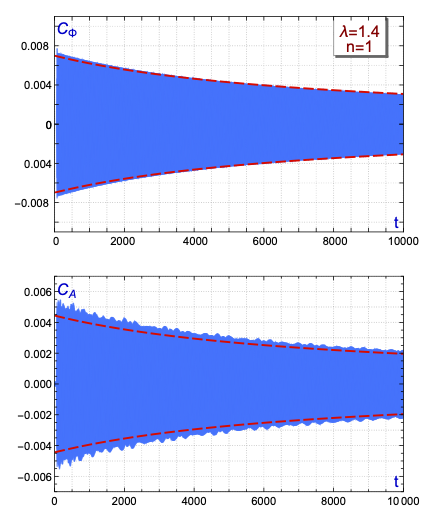}
        \caption{$\lambda = 1.4$.}
    \end{subfigure}
    \vspace{-0.0cm} 
        
    \caption{Radiation field at $r_{rad} = 50$ for different self-coupling constants $\lambda$. The upper plots account for the scalar component and the lower plots the vector component. The simulations have been performed for an initial amplitude $C_0 = 0.3$. The red dashed line represent our analytical approximation.}
    \label{Fig:Decay_Law_rad}
\end{figure}

In Figure \ref{Fig:Decay_Law_rad}, the radiation amplitudes in the scalar and vector channels at $r_{rad} = 50$ as a function of time $t$ are displayed. The dashed red lines represent the theoretical evolution of the radiation profile $(\ref{eq:decay_law_rad})$. A brief examination reveals a good agreement for the scalar component in all cases. The slight deviations in the vector channel are related to numerical precision. In order to compare the radiation amplitudes given by the formulas $(\ref{eq:rad_eta})$-$(\ref{eq:rad_xi})$ with the numerical data, we have multiplied the values from field theory by the factor $\sqrt{r_{rad}}/C_0^2$ to have a clear comparison independent of the particular position. Then, we have compared the resulting amplitudes with $C_{\phi}^{(m)}$ and $C_{A}^{(m)}$, which are given through the formulas $(\ref{eq:C_amplitudes})$. Those amplitudes and the corresponding decay rates shown in Figure \ref{Fig:Gamma_Lambda}, together with the following expressions\begin{equation}\label{eq:decay_law_rad}
C_{\phi}^{(m)}(t) = \dfrac{1}{\sqrt{(C_{\phi}^{(m)}(0))^{-2} + \Gamma^{(m)}  t}}, \hspace{1.5cm} C_{A}^{(m)}(t) = \dfrac{1}{\sqrt{(C_{A}^{(m)}(0))^{-2} + \Gamma^{(m)}  t}},
\end{equation}
have been used to predict the internal mode decay.

\subsection{The global string limit. Quasi-bound states}

\begin{figure}[ht!]
    \centering
    
    \begin{subfigure}{0.489\textwidth}
        \includegraphics[width=\linewidth]{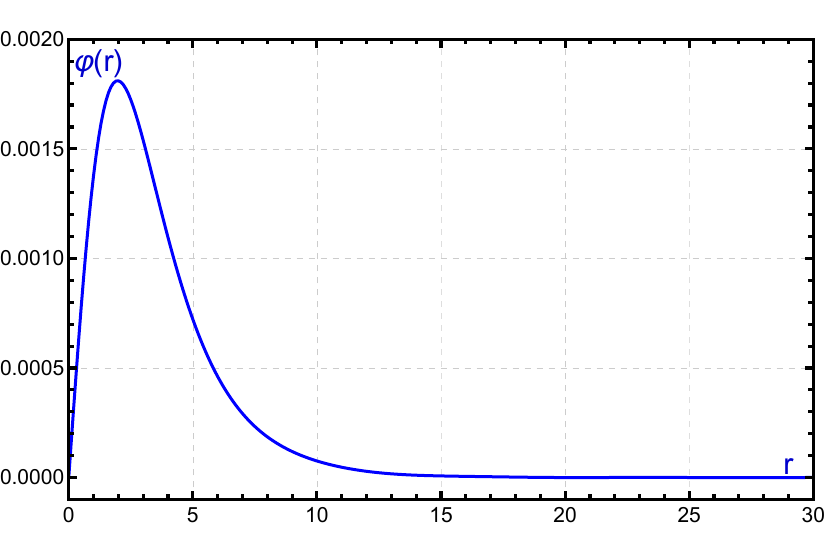}
        \caption{Scalar channel.}
       
    \end{subfigure}
    \hfill
    \begin{subfigure}{0.48\textwidth}
        \includegraphics[width=\linewidth]{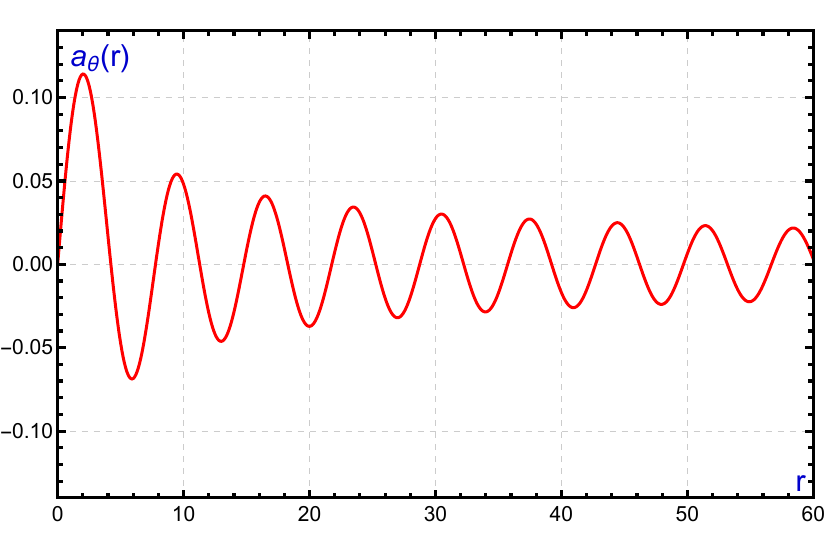}
        \caption{Vector channel.}
        
    \end{subfigure}
  
    \caption{Quasi-bound mode for the gauged vortex with $\lambda>>1$. }
    \label{Fig:GlobalMode}
\end{figure}

As we mentioned earlier, the bound state solutions of our coupled system of equations
ceases to exist when the parameter $\lambda>1.5$. The reason for this is clear, the
frequency of the bound state becomes in this case higher than the continuum for the 
vector field. This makes it impossible to have a bound state for the vector
component part of the perturbation in the analogue Schrodinger problem. However,
the frequency of the last bound state at around $\lambda=1.5$ is still below the
mass threshold for the scalar component. This suggests the possibility that
the lowest scattering states for $\lambda>1.5$ would be composed of a 
wave function similar to a bound state mode for the scalar part and
a radiative mode for the vector field part. This is indeed what one can
find in the numerical solutions presented in Figure \ref{Fig:GlobalMode}.

On the other hand, this also suggests that even though these modes 
are not truly bound states they may have a behavior that shares some similarities 
with them. In particular, it seems likely that excitations at this large coupling constants that
resemble the scalar part of these modes would have a long lifetime, comparable to the  ``bona fide" bound states described earlier. One should therefore consider
these {\it quasi-bound modes} to be qualitatively in the same family of solutions
as the genuine bound states.

We can have an intuitive understanding of the reason for the long lifetime of these
{\it quasi-bound modes} if one considers the extreme Type II regime where $\lambda >>1$. In this
limit, the scalar field has a much larger mass than the vector field and the background
vortex configuration resembles a global vortex core for $r<m_A^{-1}$. It is therefore
reasonable to expect that the scalar field excitations of the global string could
be well approximated by bound states of the complete system. 

We have investigated this idea by initializing our numerical evolution for 
a purely global string bound state in our Abelian-Higgs model with several values $\lambda >1.5 $. The 
results indicate that indeed the system bahaves like a global string for a long time
where the scalar core oscillations produced massive radiation much in the same
way as it was previously observed in \cite{BlancoPillado2021}. Furthermore, we also note the presence
of a small amount of massive vector radiation but this does not have a big effect
on the system. Presumably the coupling between the oscillating scalar core and
the vector scattering states is low enough to allow for this possibility.

\vspace{0.4cm}

\section{Concluding remarks}\label{Sec:5}

In this paper, we have analyzed in detail the evolution of a vortex excited by the lowest internal bound mode. This discrete mode exhibits rotational symmetry, enabling analytical studies based on perturbation theory. Through this approach, we have demonstrated that vortices excited by these shape modes emit radiation with radial symmetry at a frequency twice that of the shape mode, owing to quadratic nonlinear terms in the field equations. Additionally, we have provided an iterative procedure to analytically identify the decay of the shape mode amplitude, which follows an inverse square law, similar to that found in the case of the kink \cite{Manton1997} and the global vortex \cite{BlancoPillado2021}. We have performed numerical simulations that show very good agreement with our analytical predictions.

Depending of the selfcoupling values we have found different regimes. For $0.282 \leq \lambda\leq 1.5$ the excited vortex is able to decay by emitting radiation in both vector and scalar channels. For $\lambda\leq 0.28$ the vortex emits only though the scalar channel since the corresponding radiation frequency is below the vector mass threshold. Finally, for $\lambda\geq 1.5$ there are no ``proper" bound modes. However, as we have argued, if $\lambda$ is large enough the local vortex resembles a global vortex at distances $r< m_A^{-1}$. In this regime, the scalar component profiles approach the global vortex bound modes and the excited configuration is able to store energy for large times.

There are two natural extensions of these results: the study of the decay of excited vortices with $n>1$ and the decay of excited local strings in $3+1$ dimensions. In both cases, the richer spectral structure associated to static configurations requires a detailed analysis. Both lines of research are  currently under investigation.

\section*{Acknowledgments}

We would like to thank Nick Manton for his comments and fruitful discussions.
This research was supported in part by Spanish MCIN with funding from European Union NextGenerationEU (PRTRC17.I1) and Consejeria de Educacion from JCyL through QCAYLE project, as well as MCIN project PID2020-113406GB-I00. D. M. C. and S. N. O. acknowledge financial support from the European Social Fund, the Operational Programme of Junta de
Castilla y Leon and the regional Ministry of Education. J.J.B.-P. has been supported in part by the PID2021-
53123703NB-C21 grant funded by MCIN/ AEI /10.13039/501100011033/ and by ERDF;`` A way
of making Europe”; the Basque Government grant (IT-1628-22) and the Basque Foundation
for Science (IKERBASQUE).

\appendix

\section{Spectral structure computation}\label{Sec:Appendix2}

The eigensystem $(\ref{Eq:SecondOrderOperator})$ has been discretized using the following second order finite difference scheme
\begin{eqnarray}
- \dfrac{\varphi_{n;j}^{(i + 1)} - 2 \varphi_{n;j}^{(i)} + \varphi_{n;j}^{(i - 1)}}{(\Delta r)^2}  - \dfrac{\varphi_{n;j}^{(i + 1)} - \varphi_{n;j}^{(i - 1)}}{2i(\Delta r)^2}
 + \left[\dfrac{3}{2}\lambda f_n(i \Delta r)^2 - \dfrac{\lambda}{2} + \dfrac{n^2}{i^2(\Delta r)^2}\right]\varphi_{n;j}^{(i)}  && \nonumber \\ 
& & \hspace{-7.5cm} - \dfrac{n^{2}\beta_{n}(i \Delta r)}{i^2(\Delta r)^{2}}\left(2 - \beta_{n}(i \Delta r) \right)\varphi_{n;j}^{(i)}  - \dfrac{2 n f_{n}(i\Delta r)}{i\Delta r}\left(1 - \beta_{n}(i\Delta r)\right)a_{\theta,n;j}^{(i)} = \omega_{n,j}^2\varphi_{n;j}^{(i)},
\end{eqnarray}
and
\begin{eqnarray}
- \dfrac{a_{\theta,n;j}^{(i + 1)} - 2 a_{\theta,n;j}^{(i)} + a_{\theta,n;j}^{(i - 1)}}{(\Delta r)^2}  - \dfrac{a_{\theta,n;j}^{(i + 1)} - a_{\theta,n;j}^{(i - 1)}}{2i(\Delta r)^2} + \left[f_{n}(i\Delta r)^2 + \dfrac{1}{i^2(\Delta r)^2}\right]a_{\theta,n;j}^{(i)} - \dfrac{2 n f_{n}(i\Delta r)}{i\Delta r}\left(1 - \beta_{n}(i\Delta r)\right)\varphi_{n;j}^{(i)} = \omega_{n;j}^2 a_{\theta,n;j}^{(i)},
\end{eqnarray}
where $\Delta r = L/N$ and $i = 0, 1, \dots, N$, being $N$ the number of points in the mesh. We have denoted by the upper index $i$ the evaluation of the eigenfunction on $r_i = i \Delta r$, and by the lower index $j$ and $n$ the $j$-th eigenfunction appearing at vorticity $n$. Following \cite{AlonsoIzquierdo2016}, we have chosen the boundary conditions
\begin{eqnarray}
\hspace{-0.65cm} -\dfrac{4}{3}\dfrac{\varphi_{n;j}^{(1)} - \varphi_{n;j}^{(0)}}{(\Delta r)^2} + \left[\dfrac{3}{2}\lambda f_n(\Delta r)^2 - \dfrac{\lambda}{2} + \dfrac{n^2}{(\Delta r)^2}\right]\varphi_{n;j}^{(0)} && \hspace{-0.58cm} - \dfrac{n^{2}\beta_{n}(\Delta r)}{(\Delta r)^{2}}\left(2 - \beta_{n}(\Delta r) \right)\varphi_{n;j}^{(0)}\nonumber \\ 
& & \hspace{0.35cm} - \dfrac{2 n f_{n}(\Delta r)}{\Delta r}\left(1 - \beta_{n}(\Delta r)\right)a_{\theta,n;j}^{(0)} = \omega_{n,j}^2\varphi_{n;j}^{(0)},
\end{eqnarray}
and
\begin{eqnarray}
-\dfrac{4}{3}\dfrac{a_{\theta,n;j}^{(1)} - a_{\theta,n;j}^{(0)}}{(\Delta r)^2} + \left[f_{n}(\Delta r)^2 + \dfrac{1}{(\Delta r)^2}\right]a_{\theta,n;j}^{(0)} - \dfrac{2 n f_{n}(\Delta r)}{\Delta r}\left(1 - \beta_{n}(\Delta r)\right)\varphi_{n;j}^{(0)} = \omega_{n;j}^2 a_{\theta,n;j}^{(0)},
\end{eqnarray}
at the origin, and $\varphi_{n;j}^{(N)} = a_{\theta,n;j}^{(N)} = 0$ at $r = L$. Finally, the resulting $N \times N$ matrix has been diagonalized using two different algorithms from the scipy.sparse.linalg library in Python and Armadillo in C++.

\end{document}